\documentclass[modern]{aastex701}

\usepackage{amsmath}

\newcommand{\catapult}{TARS}

\begin{document}

\title{Torqued Accelerator using Radiation from the Sun (TARS) for Interstellar Payloads}

\author[orcid=0000-0002-4365-7366,sname='Kipping']{David Kipping}
\affiliation{Columbia University, 550 W 120th Street, New York NY 10027}
\email[show]{dkipping@astro.columbia.edu}  

\author[sname=Lampo]{Kathryn Lampo}
\affiliation{Columbia University, 550 W 120th Street, New York NY 10027}
\email{kel2169@columbia.edu}

\begin{abstract}
The concept of exploring space using solar power is energetically appealing, but interstellar solar sails typically require extremely low areal densities (${\sim}$\,g/m$^2$). This work explores an alternative approach: storing solar energy as rotational kinetic energy, which is later released to propel a microprobe beyond the solar system. The proposed Torqued Accelerator using Radiation from the Sun (TARS) consists of two thin surfaces with contrasting albedos that gradually spins up over weeks to months while in a sub-Keplerian ``quasite'' orbit around the Sun. Though constrained by material strengths, careful design allows a phone-sized payload to reach interstellar velocities in less than a year, using commercially available materials (e.g. CNT sheets). The entire system spans tens of meters and weighs of order of a kilogram. Whilst there is no theoretical limit to the achievable speeds, practical designs grow exponentially in size as velocity targets increase, making interstellar flight feasible but relativistic speeds implausible. Several strategies, including the use of graphene sheets, gravity assists, the Oberth effect, and electrostatic confinement, could further maximise velocity. TARS is an attractive light sail technology when high-powered directed energy systems are impractical, offering a potentially low-cost solution for deploying small, sub-relativistic interstellar probes.
\end{abstract}



\section{Introduction}
\label{sec:intro}

Light sails have long been recognised as a potential means of exploring our
solar system and beyond, powered either by natural solar radiation or directed
energy systems \citep{zander:1964,forward:1984,marx:1966,redding:1967,
moeckel:1972,weiss:1979}. An intrinsic benefit of light sails is that the fuel
needed for propulsion need not be carried onboard the spacecraft, freeing the
vessel from the so-called ``tyranny of the rocket equation''
\citep{tsiolkovsky:1968}. On the other hand, this advantage is ostensibly
intertwined with a disadvantage - radiation pressure always acts in a
direction pointing radially away from the energy source \citep{lebedev:1901,
nichols:1903}. Consequently, the spacecraft will have an acceleration vector
pointing \textit{away} from the source (barring other forces), ultimately
leading to a drop off in radiation pressure.

Certainly light sails can move towards the light source, thereby increasing
their incident radiation pressure, but they will be decelerating during such
motion and will eventually reverse and accelerate away. An exception to this
can occur for solar sails in orbit of the Sun (or indeed some other star),
where they can use angled reflectors to induce forces tangential to their
orbital motion, thus allowing them to transfer to inner (or outer) orbits 
\citep{powers:2001}.

Depending upon such transfer orbits for an interstellar light sail is not
ideal, as the transfer time is of order of the longest orbital period of the
two orbits \citep{powers:2001}, thus taking centuries to reach the Kuiper belt
for example (and technically infinite time to reach interstellar space).
Another option might be to simply direct one's solar sail outwards, but the
Sun's gravitational force will usually dwarf that of radiation pressure unless
the sail is exceptionally light.

To see this, consider the simple case of a solar sail at rest at some radial
distance $r_i$ from the Sun. The inward acceleration due to the Sun's gravity
is $G M_{\odot}/r_i^2$, whereas the idealised outward radiation force is
$L_{\odot}/(4 \pi c \Sigma r_i^2)$ (if normal to the Sun). Accordingly, an
interstellar solar sail would require a mass-per-unit-area of
$\Sigma \leq L_{\odot}/(4\pi c G M_{\odot}) = 0.77$\,g\,m$^{-2}$ for direct
radial escape.

The problem of how to get a light sail to leave our solar system is
well-recognised, and much research effort has been focussed on the idea of
directed energy systems \citep{marx:1966,redding:1967,moeckel:1972,
benford:2003,lubin:2016}. By increasing the incident flux upon the light sail
versus that caused by the Sun, the acceleration of the sail can be enhanced
as required. However, such a proposal invites new challenges, such as (but not
limited to) thermal management \citep{jin:2022}, stability within the beam 
\citep{manchester:2017,srivastava:2019,rafat:2022} and actual production and
delivery of the input energy. For example, \textit{Breakthrough Starshot}
envisages a kilometre-scale, ground-based ${\sim}$100\,GW coherent phased-array
laser \citep{worden:2021}. These challenges and others have led some to
criticise the feasibility of such a system \citep{katz:2021} and thus motivate
us to re-visit the case of a purely Sun-driven sail.

In this work, we introduce and study a new concept which does not rely on
directed energy as the primary means of propulsion. We dub this concept as a
Torque Accelerator using Radiation from the Sun (TARS) in what follows. In
Section~\ref{sec:concept}, we outline the concept. In Sections~\ref{sec:fluxes}
and \ref{sec:forces}, we calculate the fluxes and forces experienced by
\catapult. In Section~\ref{sec:spinup}, the spin-up behaviour of \catapult\ is
explored, following by a discussion of the nuances of its orbit in
Section~\ref{sec:orbit}. The achievable velocity of \catapult\ is discussed in
Section~\ref{sec:release}, as well as the impact of payload release on the
system itself in Section~\ref{sec:backreaction}. Section~\ref{sec:uniform}
introduces a specific and simplistic design realisation based upon a uniform
ribbon, which we then refine in Section~\ref{sec:tapered} to a more useful
design. Finally, we conclude in Section~\ref{sec:discussion}, commenting on the
possible applications of such a system.

\section{Concept}
\label{sec:concept}

The problem with solar sails is that they tend to move away from the source,
thus diminishing their incident flux. To this end, let us start by seeking a
solution which can harvest solar radiation without linear translation of the
vehicle, at least not initially. The idea is to store solar energy up in
some other form, essentially a battery, and then transfer said stored energy
into linear kinetic energy once the battery is charged. In this way, one can
benefit from the high incident flux found closer to a star.

A basic question is - what form might this battery take? Although several
options could be considered, in this paper we focus on an essentially
flywheel-like battery. This system is advantageous since it can be charged up
directly from solar radiation and then easily transferred to linear kinetic
energy. Indeed, it is this transfer from rotational to linear kinetic energy
that defines one of the primary safety concerns with flywheels on Earth
\citep{starbuck:2009}.

\begin{figure*}
\begin{center}
\includegraphics[width=\columnwidth,angle=0,clip=true]{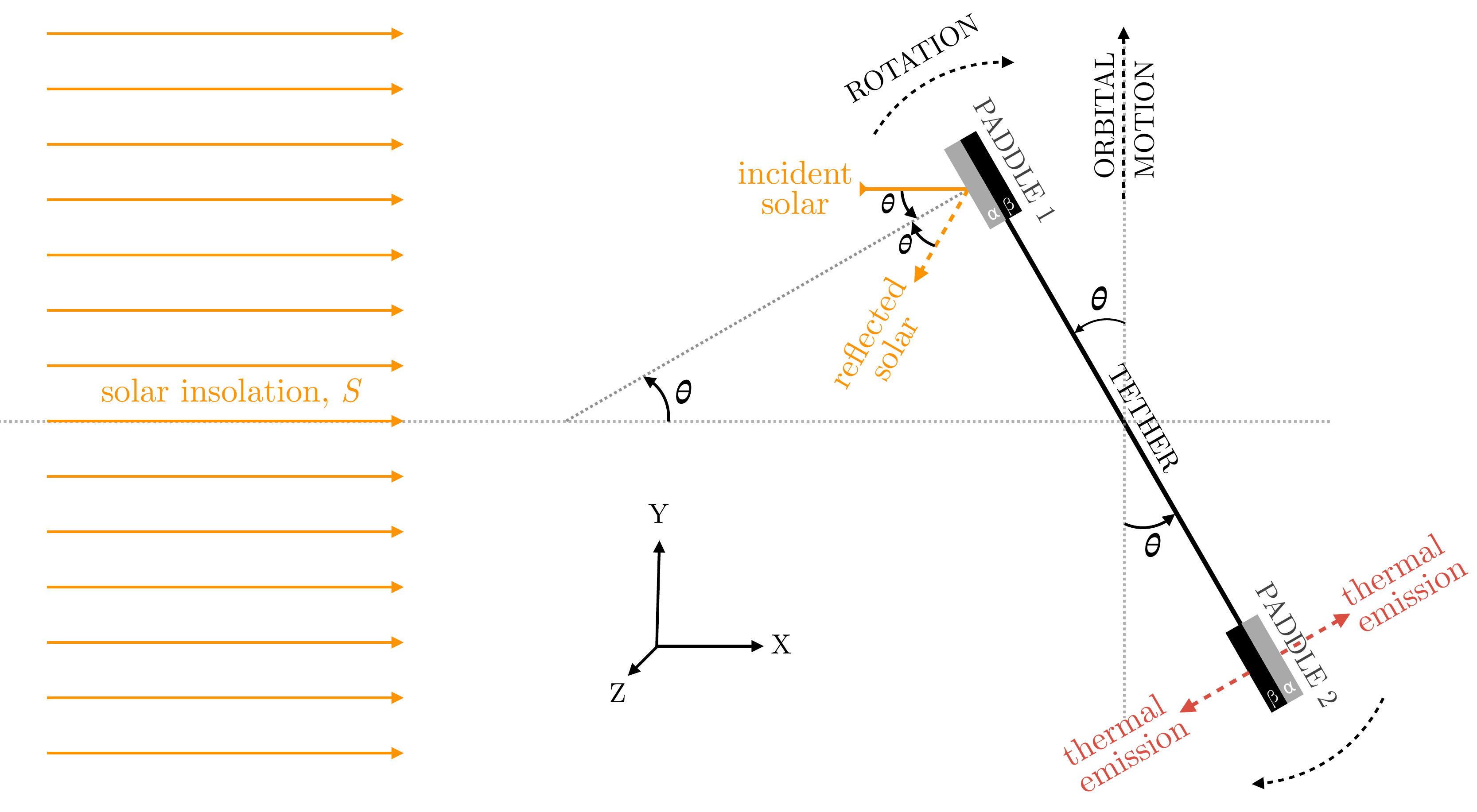}
\caption{
A simplified version of the \catapult\ system. Here, the system comprises of
one tether and two paddles, which together are orbiting around the Sun, with an
instantaneous velocity vector along the $\hat{Y}$-axis. Incident solar
radiation is largely reflected by the $\alpha$-surface (the reflective
surface) of the paddles, but largely absorbed by the $\beta$-surface. This
leads to a radiation pressure torque that gradually spins up \catapult. Note
that both paddles experience both reflection and emission; we only show one of
each for the sake of visual clarity in the above.
}
\label{fig:geometry}
\end{center}
\end{figure*}

The concept is illustrated by considering two light sails attached to one
another with a tether, as depicted in Figure~\ref{fig:geometry}. Each light
sail is identical, with one side coated with a reflective surface and the
opposite side coated with a non-reflective surface. The two light sails do not
face the same direction, but rather one is rotated 180 degrees around. In this
way, the combined system will feel both a linear translational radiation
pressure outward, as well as a torque when exposed to approximately plane
parallel radiation. The system is similar to the familiar Crookes radiometer
toy \citep{worrall:1982}.

The applied torque is utilised to spin-up \catapult\ until the tether
approaches break-up tension. At this point, one (or both) sails are detached
(or a sail section) and will head off at high speed tangential to the final
rotational motion. The light sail(s) will then continue to enjoy thrust from
solar radiation in what follows, but crucially the initial high speed
provides sufficient momentum to escape our solar system. The concept is
attractive since it only involves two light sails and a tether, and is powered
by the Sun. In practice, one might consider an initial spin-up phase with
directed energy (but far less than 100\,GW) or micro-thrusters, since 
\catapult\ is more stable once rotation is established.

The outlined concept is yet to address the issue of the linear push from solar
radiation. We propose that this radially outward force will in general be
less than the solar gravitational force in, and thus \catapult\ is still
gravitationally bound to the star. However, its orbit will trail that of
the Earth if placed at 1\,AU, since the radiation pressure effectively
reduces the gravitational mass of the Sun \citep{keze:2009}, and thus the
orbital speed of \catapult. This kind of artificial orbit, dubbed a
``quasite'' in \citet{quasite:2019}, can be placed at any location in our
solar system barring gravitational perturbations from nearby planets. In
most realistic cases the quasite effect will be fairly small (see
Section~\ref{sec:tapered}), since although the sails will experience
significant radiation pressure, the combined system including the much heavier
tether will lead to a higher areal density.

The goals of this paper are not to describe an in-depth engineering blueprint,
but rather just outline the concept along with some relevant calculations
concerning the theoretical performance. It is not claimed that this system is
definitively plausible, merely that it deserves exposition given the enormous
challenge and interest in interstellar flight. Although some obvious concerns
about feasibility are addressed later, this paper is not intended as a
feasibility study of such a system either.

\section{Fluxes}
\label{sec:fluxes}

\begin{figure*}
\begin{center}
\includegraphics[width=\columnwidth,angle=0,clip=true]{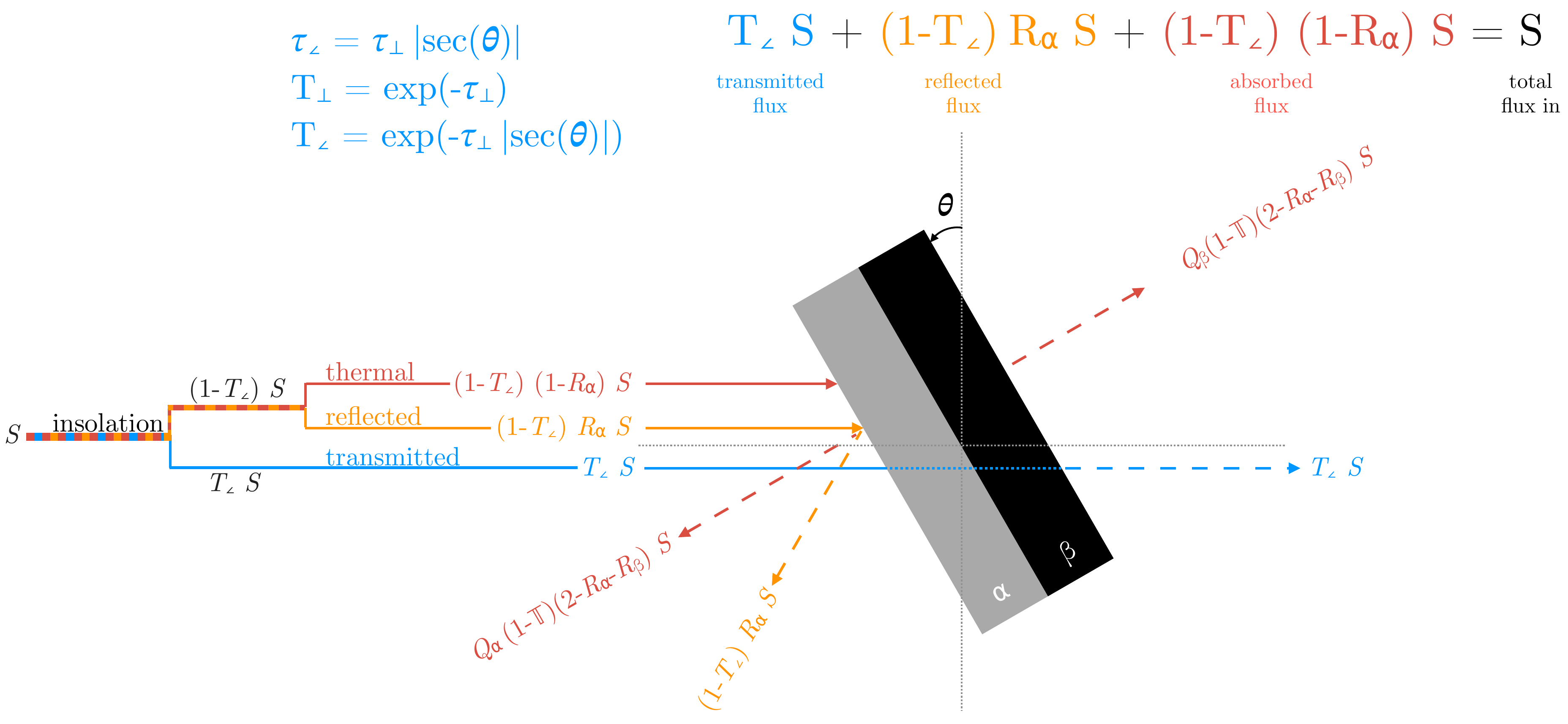}
\caption{
Break-down of fluxes incident (coloured solid) and emitted (coloured dashed)
by each surface of one of \catapult' paddles. The symbol $T$ is adopted for
transmission, $R$ for reflectivity and $Q_{\alpha}$ to describe heat transport.
The term $\mathbb{T}$ is an effective transmission term defined later in
Equation~(\ref{eqn:avgthermalpower}).
}
\label{fig:fluxes}
\end{center}
\end{figure*}

Starting from Figure~\ref{fig:geometry}, we begin by calculating the
incident/emitted fluxes for each paddle surface. In this section, it is
initially assumed that the tether does not serve as a substantial radiative
surface and can be ignored here, thus corresponding to a narrow tether in
projection. Recall that flux is power per unit area, and thus the area of the
paddles does not in fact affect the flux calculation, but will enter later when
power is calculated. A zoom-in of a paddle showing the various fluxes is
illustrated in Figure~\ref{fig:fluxes} to guide the reader, where it is assumed
that incident flux can either be i) transmitted through the paddle (blue
lines), ii) reflected (orange lines), or iii) thermally absorbed/emitted (red).

In Figure~\ref{fig:fluxes}, incident/absorbed (i.e. incoming) fluxes are shown
with solid coloured lines, whereas reflected/emitted (i.e. outgoing) fluxes are
shown with dashed coloured lines. The figure denotes a single instant in time,
which is certainly reasonable for the reflecting light rays, where it can be
safely assumed that the process of reflection occurs in a much shorter timescale
than any meaningful motion of the \catapult. However, the thermal absorption
and then subsequent re-emission cannot be trivially assumed to behave as such,
due to thermal lags. To address this, let us calculate the phase-averaged
thermal absorption and set this equal to the thermal emission at any one moment
in time. In other words, the paddles are assumed to be in thermal equilibrium
averaged over their rotation.

To make progress, it is first noted that the instantaneous incident fluxes for
phase angles of $-\pi/2<\theta<\pi/2$ equals\footnote{for the other angle set, the
equation is the same except $\alpha \to \beta$}:

\begin{align}
S &= \underbrace{T_{\angle}(\theta) S}_{\mathrm{transmitted}} + \underbrace{\big(1-T_{\angle}(\theta)\big) R_{\alpha} S}_{\mathrm{reflected}} + \underbrace{\big(1-T_{\angle}(\theta)\big) \big(1-R_{\alpha}\big) S}_{\mathrm{absorbed}}
\end{align}

where it has been assumed (for simplicity) that the albedo of the $\alpha$
surface, $R_{\alpha}$, is the same for all $\theta$ here, and hence represents
an ideal specular surface. The absorbed flux is highlighted as the final term
here. The thermal power absorbed equals this flux multiplied by the projected
area. Assuming the limiting case of a thin sail, which is certainly desired
here, the absorbing area is dominated by the projected area of the paddle's
largest surface. If the surface has an area $A$, then the projected area is
$A |\cos\theta|$. Hence, the phase-average absorbed power equals

\begin{align}
\overline{P_{\mathrm{absorbed}}} &= \frac{1}{\pi} \int_{-\pi/2}^{\pi/2} A |\cos\theta| \big(1-T_{\angle}(\theta)\big) \big(1-R_{\alpha}\big) S\,\mathrm{d}\theta \nonumber\\
\qquad& + \frac{1}{\pi} \int_{\pi/2}^{3\pi/2} A |\cos\theta| \big(1-T_{\angle}(\theta)\big) \big(1-R_{\beta}\big) S\,\mathrm{d}\theta.
\label{eqn:Pabsorbed1}
\end{align}

To make progress, it is necessary to define the transmission function,
$T_{\angle}(\theta)$. Let us define the opacity of the paddle for normal
incident light as $\tau_{\perp}$. Accordingly, the transmission through the
paddle for normal incident light would be $T(\theta=0)=\exp(-\tau_{\perp})$.
The opacity will be proportional to the path length through the paddle, which
scales as $\tau_{\perp}|\sec\theta|$. The assumption here is effectively that
the thickness layer of the $\alpha$ and $\beta$ surfaces are comparable, such
that even if they have different opacities, the net effect will still scale
with $\sec\theta$. Presumably, the opacity should be capped at some upper
limit occurring at $\theta=\pm\pi/2$ (when edge-on), since the paddle does not
have infinite width. However, both the fact that our paddle is thin and that
the transmission here will practically be negligible anyway, means this may be
ignored to simplify the integral. Accordingly, Equation~(\ref{eqn:Pabsorbed1})
becomes

\begin{align}
\overline{P_{\mathrm{absorbed}}} &= \frac{A \big(1-R_{\alpha}\big) S}{\pi} \int_{-\pi/2}^{\pi/2} |\cos\theta| \big(1-\exp(-\tau_{\perp}|\sec\theta|)\big)\,\mathrm{d}\theta \nonumber\\
&+ \frac{A \big(1-R_{\beta}\big) S}{\pi} \int_{\pi/2}^{3\pi/2} |\cos\theta| \big(1-\exp(-\tau_{\perp}|\sec\theta|)\big)\,\mathrm{d}\theta.
\end{align}

It was not possible to find a closed-form solution to the above, but after
numerically integrating along a grid of $\tau_{\perp}$ values, it was found
that the integral is well-approximated by $2 -(4/3)e^{-\tau_{\perp}}$, giving

\begin{align}
\overline{P_{\mathrm{emitted}}} = \overline{P_{\mathrm{absorbed}}} \simeq A \big(2-R_{\alpha}-R_{\beta}\big) S \underbrace{\Big( \frac{2 - \tfrac{4}{3} \exp(-\tau_{\perp})}{\pi} \Big)}_{=\mathbb{T}}.
\label{eqn:avgthermalpower}
\end{align}

where the substitution $\mathbb{T}$ absorbs the phase-averaged transmission
effects. The above is suitable to within 1\% accuracy for all
$\tau_{\perp}{>}1.34$, which corresponds to transmissions of normal incident
light of $T_{\perp}{<}26$\%. In general, a desirable paddle will have a
lower transmission than this limit and thus this formula is adopted in what
follows.

Power is not necessarily emitted equally from both sides of the paddle, due to
the different materials in use. To parametrise this, a fraction $Q_{\alpha}$ of
$\overline{P_{\mathrm{emitted}}}$ is emitted from the $\alpha$-surface, and
thus a fraction $Q_{\beta}\equiv(1-Q_{\alpha})$ is emitted from the
$\beta$-surface. In line with the thin-paddle approximation, negligible
emission comes from the edges of the paddle.

\section{Forces}
\label{sec:forces}

Equipped with the power incident/emitted off each surface, it is
straight-forward to calculate the forces at this point. One simply needs to
normalise power by the speed of light.

Splitting the forces into normal and lateral components, one may sum the
various terms in the case of an ray incident upon the $\alpha$-surface of:

\begin{align}
F_{\perp,\alpha} &= \underbrace{(A/c) \cos^2\theta(1-T_{\angle})(1-R_{\alpha}) S}_{\mathrm{inc.\,\,thml.\,\,}\alpha} \nonumber\\
\qquad& + \underbrace{(A/c) Q_{\alpha} \mathbb{T} (2-R_{\alpha}-R_{\beta}) S}_{\mathrm{em.\,\,thml.\,\,}\alpha} \nonumber\\
\qquad& - \underbrace{(A/c) Q_{\beta} \mathbb{T} (2-R_{\alpha}-R_{\beta}) S}_{\mathrm{em.\,\,thml.\,\,}\beta} \nonumber\\
\qquad& + \underbrace{(A/c) \cos^2\theta(1-T_{\angle}) R_{\alpha} S}_{\mathrm{inc.\,\,refl.\,\,}\alpha} \nonumber\\
\qquad& + \underbrace{(A/c) \cos^2\theta(1-T_{\angle}) R_{\alpha} S}_{\mathrm{em.\,\,refl.\,\,}\alpha}.
\label{eqn:Falpha1}
\end{align}

where the direction here is chosen such that a net positive force acts upon
the illuminated paddle. The $\beta$-surface is effectively the same expression,
just swapping the $\alpha$ and $\beta$ terms over.

\begin{figure*}
\begin{center}
\includegraphics[width=\columnwidth,angle=0,clip=true]{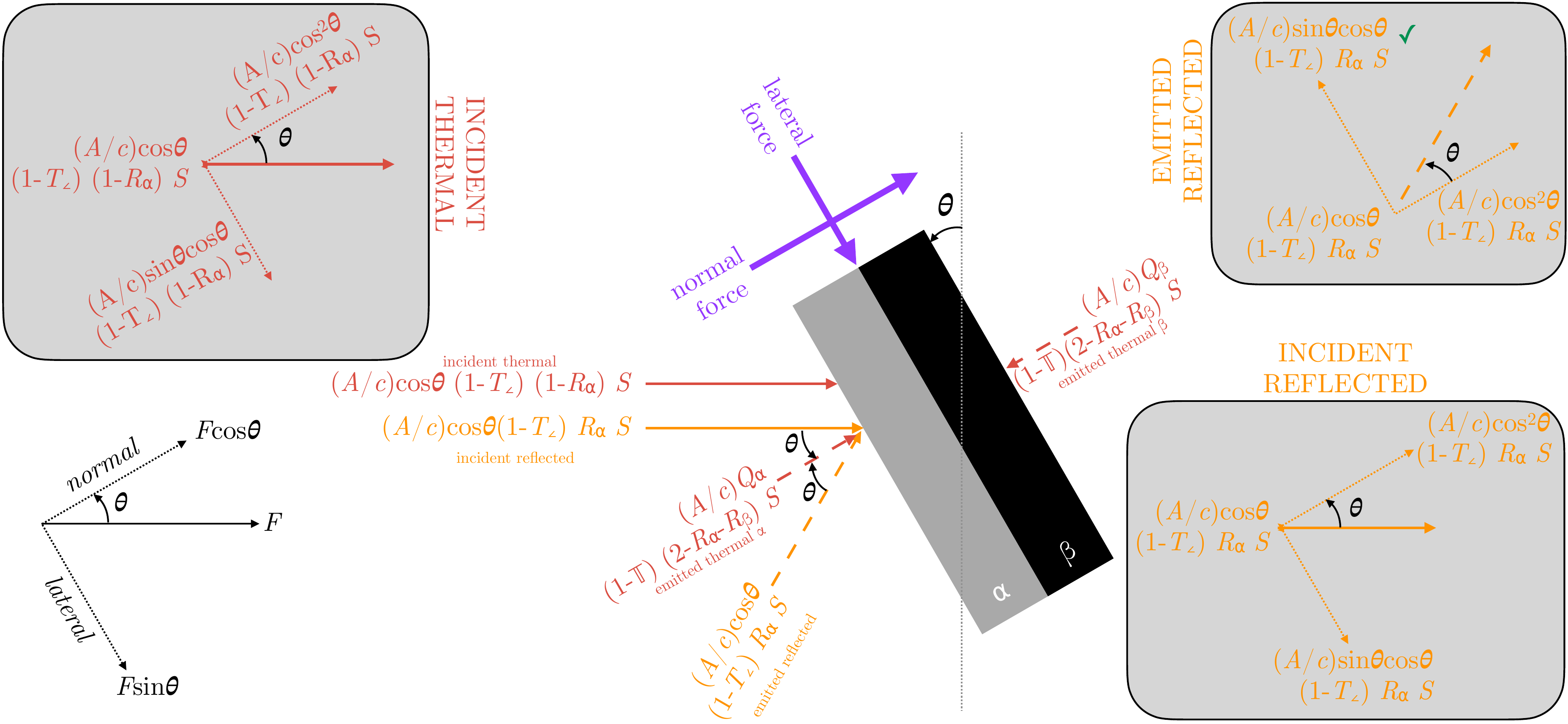}
\caption{
Break-down of the various forces acting upon a paddle when the $\alpha$-side is
Sun-incident. The spin-up of \catapult\ is governed by the normal forces,
as well as producing the quasite effect on the structure. The inset boxes show
the force components more clearly for the incident thermal (top-left), 
emitted reflected (top-right) and incident reflected (bottom-right) cases.
}
\label{fig:forces}
\end{center}
\end{figure*}

Figure~\ref{fig:forces} shows the various forces vectors and their components.
All of the forces are instantaneous forces, that is they depend on the phase
angle $\theta$, with the exception of the thermal emission components. For
these, it is assumed that the thermal response time is slower than the rotation
rate and thus it the temperature of the sail is not rapidly bouncing up and
down in phase with the spin, but rather is essentially stable over the
rotational timescale. In reality, one might expect some small damped
oscillation in temperature in phase with the rotation, but this assumption
greatly simplifies our subsequent analysis.

Equation~(\ref{eqn:Falpha1}) may now be simplified to

\begin{align}
F_{\perp,\alpha} &= S (A/c) (1+R_{\alpha}) \cos^2\theta \big( 1-\exp(-\tau_{\perp}|\sec\theta|) \big) \nonumber\\
\qquad&- S (A/c) (1-2Q_{\alpha})(2-R_{\alpha}-R_{\beta}) \mathbb{T}.
\end{align}

It is briefly noted that lateral forces also exist that in any instant do not
cancel due to the asymmetric nature of the paddle albedos. However, averaged
over all phases, these forces balance to zero.

\catapult\ is designed to impose $F_{\perp,\alpha} \geq F_{\perp,\beta}$,
such that $R_{\alpha}\gg R_{\beta}$ and $Q_{\alpha}$ does not cause a
thermal back-reaction that out-pushes the reflective radiation pressure.
In this way, the force acting on one paddle (the one currently with its
$\alpha$ surface under illumination) will exceed that of the other, creating a
torque. It is this torque which is used to gradually spin-up \catapult\ to
high velocities. In what follows, then, $F_{\perp,\alpha} \geq
F_{\perp,\beta}$ is adopted as a design requirement.

\section{Spin-Up}
\label{sec:spinup}

The motion of a rod-like object in a weightless environment being torqued at
one end is calculated in \citet{singal:2017}. In that work, two scenarios
are considered, first, that of a single impulse acting on the end of a rod, and
second, that of a continuous force, $f$. The latter closely resembles our
problem except for three differences: i) \citet{singal:2017} consider a uniform
rod with moment of inertia $I=m l^2/12$, ii) the force is continuous at
all times, unlike our phase-dependent scenario, iii) averaged over all
phases, there are no net Cartesian forces acting upon the object.

Point i) will be dealt with later in Section~\ref{sec:tapered}, where the
design of the sail is explored and the corresponding moment of inertia.

For points ii) and iii), one can correspond the forces depicted in 
Figure~\ref{fig:forces} to that presented in \citet{singal:2017} as follows.
Consider a phase position $-\pi/2<\theta<\pi/2$, such that paddle 1 has its
$\alpha$-surface under illumination, whereas for paddle 2 its $\beta$-surface
is sunlit. This is the phase position shown in Figure~\ref{fig:geometry}.
Paddle 1 thus feels a normal force of $F_{\perp,\alpha}$, acting to torque
\catapult\ in a clockwise sense (as depicted in Figure~\ref{fig:geometry}), 
whereas paddle 2 feels an opposing normal force of $F_{\perp,\beta}$, acting in
the counter-clockwise sense. Since $F_{\perp,\alpha} \geq F_{\perp,\beta}$,
then there is a net-torque in the clockwise direction, $F_R$, given by

\begin{align}
F_R &= F_{\perp,\alpha} - F_{\perp,\beta}.
\end{align}

Accordingly, one may consider that both paddles are being pushed by a force of
$F_{\perp,\beta}$, but paddle 1 experiences an extra push on top of that of
$F_R$. Because both paddles are being pushed by $F_{\perp,\beta}$, the
centre-of-mass of the system will experience a force, $F_C$, of

\begin{align}
F_C &= 2F_{\perp,\beta}.
\end{align}

In phase position $\pi/2<\theta<3\pi/2$, paddle 2 now has its $\alpha$-surface
under illumination and thus experiences the greater force. Here, the
centre-of-mass still experiences a net force $F_C$, but now paddle 2
experiences the net force $F_R$. However, given the symmetry of our system,
this is equivalent to paddle 1 experiences the force still, in terms of the
resulting torques.

One may now equate this setup to that of \citet{singal:2017}. Ignoring the
phase-dependent forces for the moment, the setups are equivalent modulo that
\catapult\ experiences an additional net force on its centre-of-mass. This
force is trying to accelerate \catapult\ radially away from the Sun. As
discussed already, such an outward acceleration is problematic since it will,
if left unchecked, cause \catapult\ to recede from the Sun and thus
decrease the incident flux. The proposed solution to this problem is to invoke
the quasite scheme, which will be discussed in further detail later in
Section~\ref{sec:orbit}.

This just leaves the issue of a phase-varying force. As \catapult\ spins
up, these become of increasingly less concern. At high speed, the motion
will asymptotically approach that of the mean forces. A similar situation
is described in \citet{singal:2017} with respect to the centre-of-mass's
motion, which experiences uneven nudges initially, but at rapid rotational
rates stabilises to a fixed point in velocity-space. Accordingly, the
rotational motion of \catapult\ will follow that of \citet{singal:2017}
except their $f$ is our phase-averaged $F_R$ ($\overline{F_R}$) and the
moment of inertia will be distinct. In the limit of $\tau_{\perp}\to\infty$
(i.e. zero transmission through the paddles), one finds

\begin{align}
\lim_{\mathbb{T}\to0} \overline{F_R} &= 2 \epsilon_R \Big( \frac{A S}{c} \Big).
\label{eqn:FRavg}
\end{align}

where $\epsilon_R$ is defined as a scalar governing the quality of \catapult'
design, here set to

\begin{align}
\epsilon_R = \frac{1}{4\pi} \Big(
16Q_{\alpha}(2-R_{\alpha}-R_{\beta}) - 16
+8 (R_{\alpha}+R_{\beta}) + \pi (R_{\alpha}-R_{\beta}) \Big).
\end{align}

One might imagine a perfect system to correspond to $R_{\alpha}\to1$,
$R_{\beta}\to0$ and $Q_{\alpha} \to 1$, yielding $\epsilon_R = (8+\pi)/(4\pi) =
0.887$. However, $\epsilon_R$ takes a maximal value of $4/\pi$ for
$R_{\alpha}=R_{\beta}=0$ and $Q_{\alpha}=1$ - a perfect thermal thruster.
In the case of an ideal reflector but no thermal thrust ($R_{\alpha}\to1$,
$R_{\beta}\to0$ and $Q_{\alpha} \to 1/2$), one obtains $\epsilon_R \to 1/4$.
After initial spin-up, the angular rotation rate will now increase linearly
with respect to time \citep{singal:2017} at a rate of

\begin{align}
\dot{\omega} &= \frac{\overline{F_R} \delta}{I},
\label{eqn:dotomega}
\end{align}

where $\delta$ is the distance from where the force $F_R$ acts to the centre of
\catapult. In the case of a homogenous ribbon of material, $\delta=L/4$
(where $L$ is the end-to-end length of \catapult), $I = M L^2/12$ and
$M = 2 \Sigma A$ (since the two paddles connect together in this simplified
case), thus yielding

\begin{align}
\dot{\omega} = 3 \frac{\epsilon_R S}{c \Sigma L}.
\end{align}

If the orbit of \catapult\ was strictly circular, then $S$ is a constant,
but in general an eccentric orbit will experience time-variable insolation.
In cases where \catapult\ will takes many years to charge to full capacity, the
insolation term can be replaced with the time-averaged insolation per orbit,
$\bar{S}$. From \citet{mendez:2017}, the time-averaged insolation is

\begin{align}
\bar{S} &= S_{\oplus} \frac{1}{(a/\mathrm{AU})^2} \frac{1}{\sqrt{1-e^2}}.
\end{align}

Adopting this, one can write that the linear speed of \catapult, after a time
$t$, will be

\begin{align}
v = 3 t \Big(\frac{\epsilon_R S_{\oplus}}{c \Sigma} \Big) \frac{1}{(a/\mathrm{AU})^2} \frac{1}{\sqrt{1-e^2}}.
\label{eqn:linspeed}
\end{align}

Some example values for the time to reach $v=10$\,km/s are provided in
Table~\ref{tab:times}. The last column shows the case of a highly
eccentric orbit, such that perihelion is two Solar radii and aphelion
is 1\,AU, corresponding to $a=0.5047$\,AU and $e=0.9816$, which increases
$\bar{S}$ by a factor of 20.55. Remarkably, the charge times are quite
modest, especially for the very low areal density cases, indicating that
a sail of such extreme minimal thickness is not strictly required.

\begin{table*}
\caption{
Some example values of the time to reach a useful velocity using
Equation~(\ref{eqn:dotomega}). The term $t_1$ is the time to reach
$v=10$\,km/s for $a=1$\,AU and $e=0$ with $R_{\alpha}=1$, $R_{\beta}=0$
and $Q_{\alpha}=1/2$. The time $t_2$ is the same except assuming a
high-performance (but sub-optimal) $\epsilon_R=(8+\pi)/(4\pi)$. The
time $t_3$ is the same as $t_2$, except for an eccentric orbit
with $r_{\mathrm{peri}}=2$\,$R_{\odot}$ and $r_{\mathrm{ap}}=1$\,AU
(a 20.5-fold increase in average insolation).
} 
\centering 
\begin{tabular}{c c c c c} 
\hline\hline
Sail & $\Sigma$ [g\,m$^{-2}$] & $t_1$ & $t_2$ & $t_3$ \\ [0.5ex] 
\hline
\textit{Lightsail2} \citep{spencer:2021} & 143 & 13.3\,years & 3.8\,years & 9.5\,weeks \\
\textit{Sunjammer} \citep{eastwood:2015} & 45.5 & 4.2\,years & 1.2\,years & 3.0\,weeks \\
\textit{Breakthrough Starshot} \citep{worden:2021} & 0.2 & 6.8\,days & 1.9\,days & 2.2\,hours \\
\hline 
\end{tabular}
\label{tab:times} 
\end{table*}

\section{Orbit}
\label{sec:orbit}

An issue that has yet to be discussed is that of the outward radial force
exerted on \catapult' centre-of-mass, $F_C$, and how this affects the orbit. As
before, let us first define the phase-averaged value this force takes, as

\begin{align}
\lim_{\mathbb{T}\to0} \overline{F_C} &= 2 \epsilon_C \Big( \frac{A S}{c} \Big),
\label{eqn:FCavg}
\end{align}

where

\begin{align}
\epsilon_C = \frac{1}{2\pi} \Big(
\pi(1+R_{\beta}) - 4(2Q_{\alpha}-1)(2-R_{\alpha}-R_{\beta})
 \Big).
\end{align}

In general, one expects $\overline{F_C}>0$ corresponds to an outward radial
force away from the energy source. Curiously, for extreme choices of
$Q_{\alpha}$ the sign reverses. Such a case is likely not realisable
without a heat pump of some kind, since it corresponds to the paddle thermally
emitting almost exclusively on a single side, implying a strong temperature
gradient. The same behaviour can of course occur for conventional sails too,
if they absorb a significant amount of flux and then primarily re-radiate on
the side facing away from the Sun.

In the typical case of a positive outward force, $\overline{F_C}$ is seemingly
a major issue since it would push \catapult\ away from its current location
like a solar sail, thereby diminishing the incident radiation. To avoid this,
it is proposed to place \catapult\ in a sub-Keplerian, quasite orbit
\citep{quasite:2019}. Quasites are the middle-ground between a statite
\citep{forward:1993} and a conventional satellite. A statite has extremely low
areal density ($\sim0.77$\,g/m$^2$), such that the outward force of radiation
pressure is sufficient to balance gravitational forces. A quasite is not so
light, but still experiences sufficient radiation pressure that a meaningful
portion of the Sun's gravitational attractive force is still balanced. In this
way, it needs to orbit the Sun, like a satellite does, in order to avoid
in-fall. However, the speed of its orbit need not be Keplerian, but rather it
is sub-Keplerian.

For a conventional solar sail, \citet{keze:2009} showed that the kinematics of
a quasite sail are well-described with Keplerian motion, except that the mass
of the Sun is effectively reduced from $M_{\star} \to \tilde{M_{\star}}$, where

\begin{align}
\tilde{M_{\star}} &= M_{\star} - \frac{ \eta L_{\star} }{2 \pi c G \Sigma}.
\end{align}

In the above, $L_{\star}$ is the stellar luminosity, $\Sigma$ is the areal
density of the quasite. \citet{keze:2009} don't explicitly define $\eta$ but
describe that $\eta=1$ corresponds to a perfectly reflective sail and
$\eta=1/2$ is a perfectly absorbing sail.

To relate $\eta$ to $\overline{F_C}$, consider what $\overline{F_C}$ would
be if replaced \catapult\ with a simple solar sail orthogonal to the incident
radiation. In the case of no transmission, one finds

\begin{align}
\lim_{T_{\perp}\to0} \overline{F_C} &= \frac{A S}{c} 2 (R + Q(1-R)),
\label{eqn:sailforce}
\end{align}

where $Q$ is the ratio of heat emitted from the Sun-ward side to the
space-ward side. In the limit of a perfect reflector, what
\citet{keze:2009} call $\eta=1$, one obtains

\begin{align}
\lim_{R\to1} \lim_{T_{\perp}\to0} \overline{F_C} &= 2 \frac{A S}{c}.
\end{align}

For the perfect absorber, $\eta=1/2$, instead one finds

\begin{align}
\lim_{Q\to1/2} \lim_{R\to0} \lim_{T_{\perp}\to0} \overline{F_C} &= \frac{A S}{c}.
\end{align}

From this, one can see that $\eta$ is simply defined as

\begin{align}
\eta &= \frac{ \overline{F_C} }{ 2A S/c}.
\end{align}

Accordingly, in our case, $\eta=\epsilon_C$ and thus

\begin{align}
\tilde{M_{\star}} &= M_{\star} - \frac{ \epsilon_C L_{\star} }{2 \pi c G \Sigma}.
\label{eqn:Mtilde}
\end{align}

If \catapult\ remains bound to the Sun, then one requires that the last term does
not exceed $M_{\star}$.

\section{Payload Release}
\label{sec:release}

As \catapult\ spins up, the ends eventually approach some desired target
velocity, $v_{\mathrm{targ}}$, at which point the payload is released. Beyond
this speed, there will also be a critical velocity, $v_{\mathrm{crit}}$,
where \catapult\ tears itself apart from centrifugal forces.

In the case where \catapult\ is on a circular orbit, the moment of release
need not be precisely when $v=v_{\mathrm{targ}}$, but rather can be at a
specific orbital phase position instead. In this way, the direction of the
payload can be controlled - although it is fated to always lie in the plane of
\catapult' orbit and thus careful consideration of the orbit will be required.
Through detailed calculation, it may be possible to synchronise the orbital
phase position of the release with the moment when target velocity is reached,
thus allowing one to engineer $v_{\mathrm{targ}}$ as close to
$v_{\mathrm{crit}}$ as is dared. There will certainly be some control here
since \catapult\ may experience an initial spin-up phase and/or could be
further fine-tuned with directed energy upon the paddles.

It is instructive to consider the potential of \catapult\ to reach
interstellar escape velocity, $v_{\mathrm{esc}}$, given by

\begin{align}
v_{\mathrm{esc}} &= \sqrt{ \frac{ 2 G \tilde{M_{\star}} }{ r }},
\label{eqn:vesc}
\end{align}

where $r$ is the radial distance from the Sun. Note how the familiar
$M_{\star}$ term is replaced with $\tilde{M_{\star}}$
(Equation~\ref{eqn:Mtilde}) due to the quasite effect, which essentially
captures how the payload enjoys an outward radiation pressure during its exodus
from our solar system. In the limit of a circular orbit, the required target
velocity to escape our solar system will be

\begin{align}
\lim_{e\to0} v_{\mathrm{targ},\mathrm{req}} &= (\sqrt{2}-1) \sqrt{\frac{G \tilde{M_{\star}}}{r}},
\end{align}

and for $r=1$\,AU this demands a 12.3\,km/s rotational velocity using
\textit{Lightsail2}-like parameters with $\epsilon_C=1$. This highlights how
it is challenging, but feasible, to engineer a \catapult\ system which can go
interstellar, a topic discussed in greater depth in
Section~\ref{sec:tapered}. Even without interstellar speeds, \catapult\ is
potentially still useful for interplanetary missions, but let us briefly consider
how the system could be modified to reach interstellar space in what follows. One
obvious possibility to increase $r$, since

\begin{align}
v_{\mathrm{targ},\mathrm{req}} &\propto \frac{1}{\sqrt{r/\mathrm{AU}}}.
\end{align}

For example, placing \catapult\ at or beyond Mars' orbit would just about
achieve escape velocity with $v_{\mathrm{targ}} = 10.0$\,km/s, using
\textit{Lightsail2}-like properties. However, since
$v_{\mathrm{targ}} \propto r^{-2}$ (see Equation~\ref{eqn:linspeed}),
this more than doubles the spin-up time, although given the times in
Table~\ref{tab:times} this is perhaps perfectly acceptable. Of course,
\catapult\ would have to be first be manoeuvred to or constructed in such an
orbit, which is clearly less practical.

An alternative is for \catapult\ to have an eccentric orbit to begin with.
This obviously requires some extra energy to move the vehicle into such an
orbit, but has some considerable benefits in terms of payload speed. A wide
array of orbits could be considered here, but for the sake of demonstrating its
advantages, consider an orbit with perihelion of $20$\,$R_{\odot}$ and
aphelion 1\,AU, such that $a=0.5465$\,AU and $e=0.8298$ - which is less
extreme that achieved by NASA's \textit{Parker} probe. Releasing the payload
at perihelion, with the same parameters otherwise as used earlier, yields

\begin{align}
v_{\mathrm{targ},\mathrm{req}} &= 6.0\,\mathrm{km}/s.
\end{align}

In other words, the critical velocity is approximately halved thus greatly
reducing the engineering requirements on the system. Effectively, \catapult\
here is exploiting the ``Oberth effect’’ \citep{blanco:2019}. The kinetic energy of
\catapult\ is maximised at perihelion, and hence applying a delta-v at this
point maximises the energy gain.

For the suggested orbit, the perihelion temperature would peak at 913\,K on the
$\beta$-paddles, which although is too high for Mylar, is below the melting
point of metals such as aluminium (934\,K) and beryllium (1560\,K) which could
be polished for the reflective coatings. With this orbit,
$\bar{S} = 6\,S_{\oplus}$ and thus we still enjoy decreased charge times
compared to the nominal circular 1\,AU case. In what follows, the rest of
this study considers strictly circular orbits at 1\,AU. Although this is
not optimal for maximal speeds, it is optimal in terms of launch energetics
thereby focussing on the most viable realisation in a practical sense.

\section{Backreaction \& Recharging}
\label{sec:backreaction}

After the payload release, the angular momentum of \catapult\ will be
affected. The effect is estimated in what follows, using the simplifying
assumption of a uniform density ribbon (thin rectangular plate). Different
designs will of course yield different results, but the below provides
intuition about the expected outcomes.

Consider the system has an angular momentum of $J=I\omega$ just before release,
where $I=\tfrac{1}{12} M L^2$. For the payload release, imagine an end of the
paddle being removed of length $\Delta L$ and mass $\Delta M$, such that the
new moment of inertia becomes $I'=\tfrac{1}{12} (M-\Delta M) (L-\Delta L)^2$.

At the same time, the payload itself carries away an angular momentum
$r \times p$, where $p = v \Delta M = \omega r \Delta M$ and $r$ is the
distance of the payload's barycentre from the rotation axis, given by
$r=(L/2 - \Delta L/2)$. Thus, the new angular momentum of the system is

\begin{align}
J' &= \frac{1}{12} M L^2 \omega - r^2 \Delta M \omega,\nonumber\\
\qquad&= \frac{1}{12} M L^2 \omega - (L/2 - \Delta L/2)^2 \Delta M \omega,
\end{align}

which can be equated to the new momenta of inertia and angular velocity as

\begin{align}
\frac{1}{12} (M-\Delta M) (L-\Delta L)^2 \omega' &= \frac{1}{12} M L^2 \omega - (L/2 - \Delta L/2)^2 \Delta M \omega.
\end{align}

And thus the new angular velocity, relative to the old, can be written as

\begin{align}
\frac{\omega'}{\omega} &= \frac{1 - 3 \Delta m (1-\Delta l)^2 }{(1-\Delta m)(1-\Delta l)^2},
\end{align}

which uses the substitutions $\Delta m = \Delta M /M$ and
$\Delta l = \Delta L/L$. Since uniform density is assumed,
then $\Delta m = \Delta l$ and thus

\begin{align}
\frac{\omega'}{\omega} &= \frac{1 - 3 \Delta m (1-\Delta m)^2 }{(1-\Delta m)^3},
\end{align}

which is greater than unity for all $\Delta m>0$ and reaches a maximum value of
5 for $\Delta m \to 1/2$. Since the velocity of the ends is given by
$\omega' L'$, then the new velocity can be shown to satisfy

\begin{align}
\frac{v'}{v} &= \frac{1 - 3 \Delta m (1-\Delta m)^2 }{(1-\Delta m)^2},
\label{eqn:backreaction1}
\end{align}

which is less than unity (as perhaps our intuition might expect) for all positive
$\Delta m$, taking a limit of $5/2$ when $\Delta m \to 1/2$. For payloads of
relatively small mass, one finds

\begin{align}
\frac{v'}{v} &= 1 - \Delta m + \mathcal{O}[\Delta m^2].
\end{align}

If the payload was released close to $v_{\mathrm{crit}}$, which is independent
of $M$ or $L$ for a uniform density ribbon, then \catapult\ will slow down to a
sub-critical velocity after release. Afterwards, \catapult\ will continue to
feel a torque, although the centre-of-force has now shifted due to the
asymmetrical release at just one end. In principle then, it should recharge
back up to critical velocity and could be used to launch another payload.
Indeed, a series of payloads could be released in this way. To ensure the
rotational axis does not shift and potentially lead to unstable rotation, a
double payload could be released, one from each end (even if one of these is
a ballast payload). This modification changes
Equation~(\ref{eqn:backreaction1}) to

\begin{align}
\frac{v'}{v} &= \frac{1 - 6 \Delta m (1-\Delta m)^2 }{(1-2\Delta m)^2},
\label{eqn:backreaction2}
\end{align}

or simply

\begin{align}
\frac{v'}{v} &= 1 - 2\Delta m + \mathcal{O}[\Delta m^2].
\end{align}

This notion of a series of payload releases separated by a recharge time is
attractive since the recharge time will be a small fraction of the initial
charge time, making \catapult\ more cost-efficient by releasing multiple
probes. Indeed, using Equation~(\ref{eqn:linspeed}), the recharge time will be

\begin{align}
\Delta t &\simeq \frac{2}{3}\Delta m v_{\mathrm{crit}} \Bigg(\Big(\frac{\epsilon_R S_{\oplus}}{c \Sigma} \Big) \frac{1}{(a/\mathrm{AU})^2} \frac{1}{\sqrt{1-e^2}}\Bigg)^{-1}
\end{align}

Given the constraint that payloads must be released in the same plane, a
possible strategy is to release a series of micro-probes to the same
destination, forming a daisy-chain physically separated by
${\sim} v_{\mathrm{crit}} \Delta t$. Such a system could allow for the
low-power probes to maintain a communication line back to Earth.

\section{Uniform Ribbon}
\label{sec:uniform}

In what follows, let us consider what is arguably the simplest version of
\catapult\ - a homogenous ribbon. Here, there is no tether, the paddles extend
and mate at the midpoint. The midpoint is identified as the point where the
coatings change, from $\alpha$ to $\beta$ values. More optimal designs surely
exist, as discussed in Section~\ref{sec:tapered}, but the uniform ribbon is
attractive for simplifying the calculations in what follows.

\subsection{Critical Velocity}

First, let us derive the critical velocity at which the ribbon will fail due
to the centrifugal forces exceeding the tensile strength. The ribbon (really a
rectangular plate) is defined to have a length $L$, width $W$, thickness $t$,
density $\rho$ and tensile strength $\sigma$.

Consider an element of the ribbon of length of $\mathrm{d}x$ of the ribbon,
located a distance $x$ from the centre and rotating at an angular velocity
$\omega$. The centrifugal force experienced by this element will be

\begin{align}
\mathrm{d}F &= \rho W t \mathrm{d}x \omega^2 x.
\end{align}

The tension, $\mathcal{T}$, acting at a distance $x$ from the centre can be
found by integrating the above from position $x$ to $L/2$, such that the
tension is the sum of the centrifugal forces acting outward beyond the point
$x$:

\begin{align}
\mathcal{T}(x) &= \int_{x}^{L/2} \rho W t \omega^2 x\,\mathrm{d}x,\nonumber\\
\qquad&= \frac{1}{8} (L^2-4x^2) W t \rho \omega^2.
\end{align}

At $x=L/2$, we can see that the tension falls to zero  (i.e.
$\mathcal{T}(L/2)=0$), as expected for the boundary conditions of the problem.
At the midpoint, the structure experiences a maximum tension of

\begin{align}
\mathcal{T}(0) &= \frac{1}{8} L^2 W t \rho \omega^2.
\end{align}

The failure load is defined as the tensile strength multiplied by the
cross-sectional area, and hence here equals $\sigma W t$. Equating this to
$\mathcal{T}(0)$ and re-arranging, one obtains

\begin{align}
\omega_{\mathrm{crit}}^2 &= \frac{8\sigma}{\rho L^2},
\end{align}

and thus

\begin{align}
v_{\mathrm{crit}} &= \sqrt{2}\sqrt{\frac{\sigma}{\rho}}.
\end{align}

\subsection{Material}

A key design requirement is to maximise the specific strength i.e.
$\sigma/\rho$. Graphene is the most promising material in this regard with a
density of $2.3$\,g\,m$^{-3}$ and tensile strength of $130$\,GPa. However,
graphene is difficult to produce in large continuous sections and thus
the following considers instead carbon nanotube (CNT) sheets, with a density
of $1.6$\,g\,m$^{-3}$ and tensile strength up to $20$\,GPa. Although CNT sheets
have a specific strength five terms worse than graphene, they are a more
mature product and one which is currently available on the market.

For the thickness of \catapult, one might apply a thin optical coating
on either side of CNT sheets, of at least 10\,nm. Given that the optical
coatings will have negligible tensile strength, they dilute the overall
tensile strength. To overcome this, it is proposed to make the CNT sheets much
thicker than 10\,nm, such that the bulk properties asymptotically approach
that of pure CNT sheets. Of course, the more mass that is added, the longer the
spin-up time and thus a possible compromise is a CNT sheet of order of a micron
thickness, which is already in widespread production.

For the $\alpha$ surface, a possible choice is a thin (${\sim}10$\,nm or 35
atoms thick) coating of nanostructured silver to deliver low transmission
and high reflectance (${\sim}0.9$) due to localised surface plasmon resonances
\citep{kuzminova:2019}. Other options might include silver or beryllium with
dielectric coatings (such as SiO$_2$, TiO$_2$ or HfO$_2$) or doping the coating
\citep{atwater:2018}. Such a coating would be painted onto the CNT sheets using
processes such as physical vapour deposition \citep{mattox:1998} or chemical
vapour deposition \citep{ohring:2002}. Aluminium and lithium should be avoided
if aphelion $\lesssim20$\,$R_{\odot}$ is planned, due to their lower melting
points.

For the $\beta$ surface, a possible choice is a thin (${\sim}10$\,nm or 71
atoms thick) coating of titanium nitride \citep{patsalas:2015}. Even this thin
layer would have negligible transmission and deliver low optical reflectance
(${\sim}0.3$).

As presented, this ribbon has no payload but one could certainly imagine adding
payloads of low mass compared to the ribbon at the ends without significant
performance losses. Rather than go into the minutiae of such a proposal, let us
instead move onto an improved design which is more interesting to consider in
detail.                                           

\section{Tapered Ribbon}
\label{sec:tapered}

\subsection{Payload-Free Scenario}

To begin, consider the case of a payload-free tapered ribbon. In the
previous section, the critical angular velocity was calculated by setting
the tension at the midpoint to be equal to the failure load. However,
if one moves further along the ribbon, the tension decreases and thus
the ribbon is unnecessarily wide. Thus, one should be able to improve
performance by gradually narrowing the width of the ribbon towards the
ends - a tapered ribbon.

Let us split the ribbon up into $N$ equi-length segments from $x=0$ to $x=L/2$,
assigning each one a unique width $W_i$ and segment length $(L/2)/n$.
Imposing symmetry reflects this configuration from $x=0$ to $x=-L/2$. The
bounds around each segment follow $(i/n) (L/2)$ from $i=0$ (the midpoint) to
$i=n$ (the end). One may define a general $\mathcal{W}(x)$ piecewise function
which returns $W_i$ when $x$ corresponds to a distance (from the midpoint)
within the $i^{\mathrm{th}}$ segment.

Beginning at the midpoint, the critical angular velocity is again found by
balancing the tension and the failure load - a solution labelled as
$\omega_{\mathrm{crit},1}$. One may then move along up to $i=n$ solving the
following equation each time for $\omega$:

\begin{align}
\int_{x}^{L/2} \rho \mathcal{W}(x') t \omega^2 x'\,\mathrm{d}x' = \sigma \mathcal{W}(x) t.
\end{align}

Let us set $\omega_{\mathrm{crit},i}=\omega_{\mathrm{crit},i-1}$ and solve for
$W_i$, working backwards from $i=n$ to $i=2$ ($W_1$ is treated as a design
parameter), which yields the following recursive solution

\begin{align}
W_i &= \frac{2(n+1-i)}{2n-1} W_{i-1}.
\label{eqn:Wrecursive}
\end{align}

Accordingly, in terms of $W_1$ only, these solutions become

\begin{align}
W_i &= \frac{ \prod_{j=2}^i 2 (n+1-j) }{ (2n-1)^{i-1} } W_1.
\end{align}

Applying these solutions, one finds, for all $i\in[1,n]$:

\begin{align}
\omega_{\mathrm{crit},i} &= \frac{n}{\sqrt{2n-1}} \times \underbrace{\Bigg( 2\sqrt{2} \frac{ \sqrt{\sigma} }{ \sqrt{\rho}L }  \Bigg)}_{\text{uniform solution}}.
\label{eqn:wcritn}
\end{align}

Hence, this tapering scheme allows one to arbitrarily increase
$\omega_{\mathrm{crit}}$ (at the expense of ever larger structures).

Each segment has a width $W_i$, length $(L/2)/n$ and uniform thickness $t$.
The area of each segment is thus $A_i = W_i (L/2)/n$. Given that the assumption
of uniform density, $\rho_{\mathrm{bulk}}$, then the mass elements are
$M_i = A_i t \rho_{\mathrm{bulk}}$. As before, there is still no payload but 
a scheme for including payload is presented in what follows.

\subsection{Adding Payload}

\begin{figure*}
\begin{center}
\includegraphics[width=\columnwidth,angle=0,clip=true]{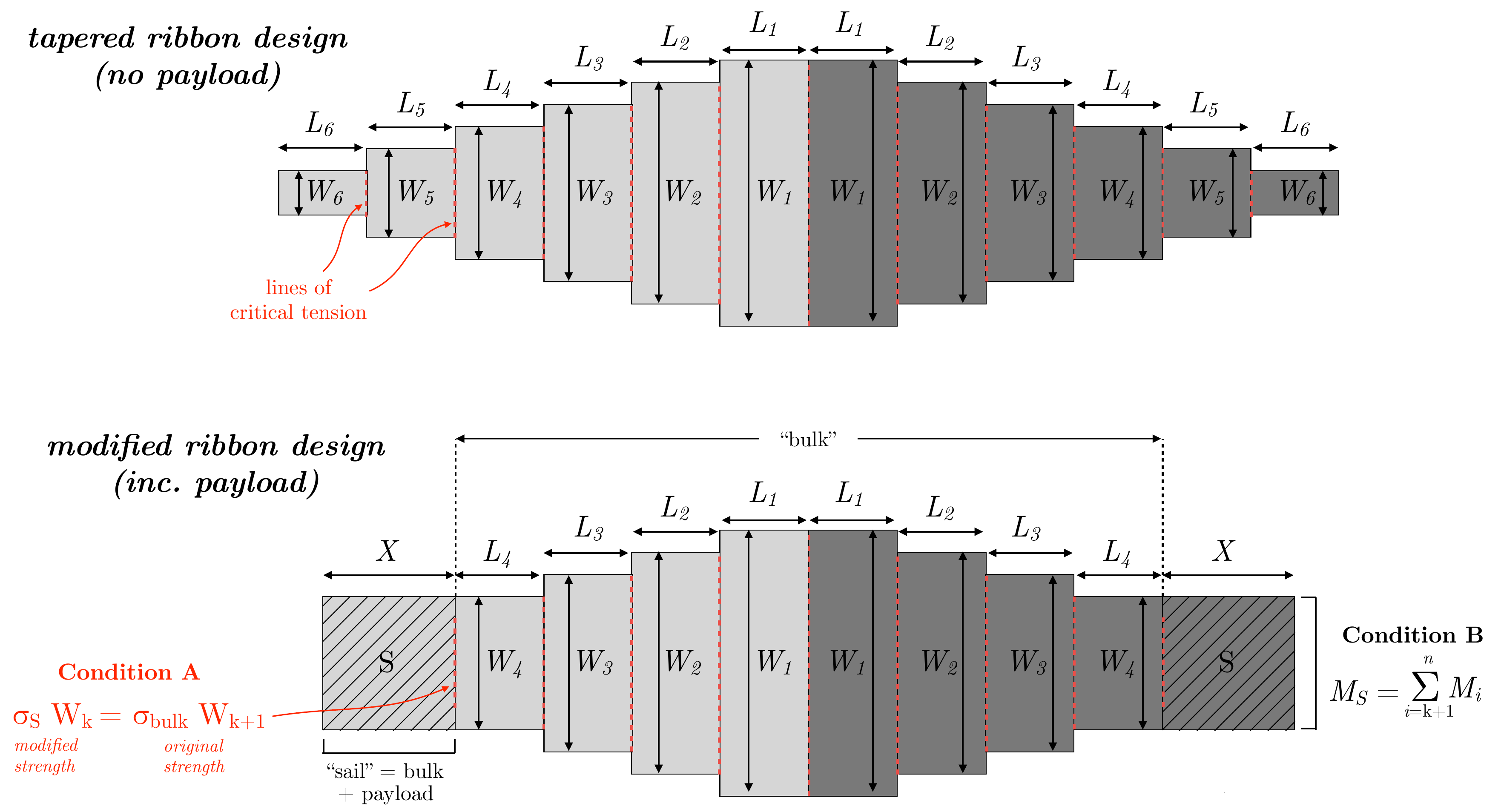}
\caption{
The tapered ribbon design of \catapult. Top: The case of a tapered
ribbon $(n=6)$ with no payload and made of a homogeneous material (the bulk).
The $\alpha$ and $\beta$ sides are shaded appropriately. Moving away from the
rotation axis, the segments become of ever smaller widths, $W_i$, since
the tension reduces and thus less tensile strength is required to resist
failure. The red lines depict points of critical tension when \catapult\
reaches $\omega_{\mathrm{crit}}$. Bottom: A modification of the above where
the $k=4$ inner segments are kept but those beyond are replaced with a sail
containing the payload. The sail has poorer tensile strength due to the payload
and thus the width is inflated to provide the necessary additional tensile
strength. Condition A is the criterion to avoid the sail breaking off before
$\omega_{\mathrm{crit}}$, and condition B conserves mass.
}
\label{fig:design}
\end{center}
\end{figure*}

Consider replacing segments $k+1$ to $n$ with a single segment that
includes the same basic materials, a core of CNT sheets with the same optical
coatings, but also includes a thin payload layer. For simplicity, the
thickness of this new section is considered to be equal to that of the rest of
\catapult\ (the ``bulk''), which means the thickness of CNT sheets (and thus
specific strength) is less in this end section to accommodate the payload. The
proposal is depicted in Figure~\ref{fig:design}.

As the payload-containing segment will ultimately be released and fly as a
sail, it will referred to as the ``sail'' segment in what follows. Due to the
diminished specific strength of the sail, to avoid failure one requires that
$W_S> W_{k+1}$ to provide sufficient tensile strength. This work chooses to set
$W_S=W_k$ to create a continuous join. This increased cross-sectional area
provides added tensile strength to compensate for the weaker specific
stength of the composite sail material. Since the thickness of the sail and
bulk are matched, then to avoid failure one requires condition A of

\begin{align}
\sigma_S W_k &= \sigma_{\mathrm{bulk}} W_{k+1},\nonumber\\
\sigma_S &= \sigma_{\mathrm{bulk}} \frac{2(n-k)}{2n-1}.
\end{align}

where the second line uses Equation~(\ref{eqn:Wrecursive}). By volumetric
mixing, $t_{\mathrm{bulk}} \sigma_S =
t_{\mathrm{payload}}\sigma_{\mathrm{payload}} +
(t_{\mathrm{bulk}} - t_{\mathrm{payload}}) \sigma_{\mathrm{bulk}}$, which
allows one to re-write condition A as:

\begin{align}
1 - \frac{t_{\mathrm{payload}}}{t_{\mathrm{bulk}}} &= \frac{2(n-k)}{2n-1}.
\end{align}

One may use this condition to calculate the mean density of the sail component,
since again by volumetric mixing one has

\begin{align}
\rho_S &= \Big(\frac{t_{\mathrm{payload}}}{t_{\mathrm{bulk}}}\Big) \rho_{\mathrm{payload}} + \Big(1-\frac{t_{\mathrm{payload}}}{t_{\mathrm{bulk}}}\Big) \rho_{\mathrm{payload}}.
\end{align}

Next, let us impose a condition B - that the mass of the sail equals the
cumulative mass of the removed sections (meaning that that \catapult's total
mass is unaffected by our modification), such that

\begin{align}
M_S &= \sum_{i=k+1}^n M_i,\nonumber\\
\rho_S W_k X &= \sum_{i=k+1}^n W_i \frac{L}{2n} \rho_{\mathrm{bulk}}.
\end{align}

This ensures the general formula for the critical angular velocity found
earlier (Equation~\ref{eqn:wcritn}) remains true. The above may be re-arranged
to give a solution for $X$ as a function of $L$, a solution that will be used
shortly to define the final dimensions of the system.

The spin-up of \catapult\ will now be governed by the moment of inertia
and centre-of-pressure location, following Equation~(\ref{eqn:dotomega}). The
centre-of-pressure distance, $\delta$, is simply the centre-of-area since
the \catapult\ has a uniform optical coating, and is thus

\begin{align}
\delta &= \frac{ X W_k d_S + \sum_{i=1}^k W_i \frac{L}{2n} d_i }{ X W_k + \sum_{i=1}^k W_i \frac{L}{2n} }
\end{align}

where $d_i = (2i-1)L/(4n)$ and $d_S = ((k L)/(2n))+(X/2)$ are the distances
of the relevant sections from the axis of rotation. The moments of inertia
are calculated using

\begin{align}
I_X &= 2\sum_{i=1}^k \int_{x=(i-1)(L/2)/n}^{i(L/2)/n} \int_{y=-W_i/2}^{W_i/2} \int_{z=-t/2}^{t/2} \rho_{\mathrm{bulk}} (y^2+z^2)\,\mathrm{d}x\,\mathrm{d}y\,\mathrm{d}z \nonumber\\
\qquad& + 2 \int_{x=k(L/2)/n}^{X+k(L/2)/n} \int_{y=-W_k/2}^{W_k/2} \int_{z=-t/2}^{t/2} \rho_S (y^2+z^2)\,\mathrm{d}x\,\mathrm{d}y\,\mathrm{d}z
\end{align}

and similarly for $I_Y$ and $I_Z$. These are particularly important for not
only the spin-up time but also designing a system that does not risk tumbling
\citep{goldstein:1980}. The axis of rotation is along the $Y$-direction and
thus one must ensure that the intermediate moment of inertia is not $I_Y$,
which in practice means designing a system such that $I_Y<I_X<I_Z$.

The velocity of the sail, as it approaches the critical angular rotational
speed, is

\begin{align}
v_{\mathrm{crit}} &= d_S \omega_{\mathrm{crit}}.
\end{align}

If released tangential to the orbital motion, the total speed of the sail will
additive to this extra speed. However, because \catapult\ is a quasite, the
orbital speed is sub-Keplerian, given by

\begin{align}
v_{\mathrm{orb}} &= \sqrt{\frac{2}{r}-\frac{1}{a}} \sqrt{G} \sqrt{M_{\star} - \frac{\epsilon_C L_{\star}}{2\pi c G \Sigma_{\mathrm{eff}}}},
\end{align}

where $\Sigma_{\mathrm{eff}}$ is the total mass of \catapult\ divided by
its area. To maximise radial escape, the sail will orient such that the
$\alpha$-side only is Sun-facing, thus modifying the quasite effect. The
escape velocity from the Sun is then given by Equation~(\ref{eqn:vesc})
but with $\epsilon_C=(1+R_{\alpha})/2$.

\subsection{Numerical Examples}

This work does not attempt a comprehensive optimisation of the design
parameters, but instead this subsection explores the effect of some of the
parameters to provide guidance on the plausible capabilities.

To set the scale of the system, this work sets $W_k=0.1$\,m as a test case,
yielding a sail with an area of $0.01$\,m$^2$ - comparable to the dimensions of
a modern smart phone. For the thickness, let us adopt a 6\,$\mu$m thick layer
of CNT sheets and a 20\,nm total thickness of optical coating. One may now
experiment with different choices of $n$ and $k$ and solve for the final system
parameters. In all cases, a circular orbit is assumed, as well as
$S=S_{\oplus}$ (i.e. an Earth-trailing orbit). Further, it is assumed that
$Q_{\alpha}=1/2$, $R_{\alpha}=0.9$ and $R_{\beta}=0.3$.

Consider first the simplest case of $n=2$ and $k=1$. In this case, one finds
$v_{\mathrm{crit}}=3.6$\,km/s for which the spin-up time is 263.9\,days.  
\catapult\ has an effective areal density of 10.9\,g/m$^2$ whereas the sail 
component is 13.1\,g\,m$^2$, comprised of a payload thickness of 2007\,nm. This  
configuration is unstable to tumbling and also lacks the ability to go to
interstellar, with $v_{\mathrm{orb}}+v_{\mathrm{crit}}=32.0$\,km/s, $7.8$\,km/s 
shy of the escape speed. With a length of 0.8\,m and a $W_1$ width of
$0.1$\,m, the structure has a high aspect ratio. 

\begin{figure}
\begin{center}
\includegraphics[width=12.0cm,angle=0,clip=true]{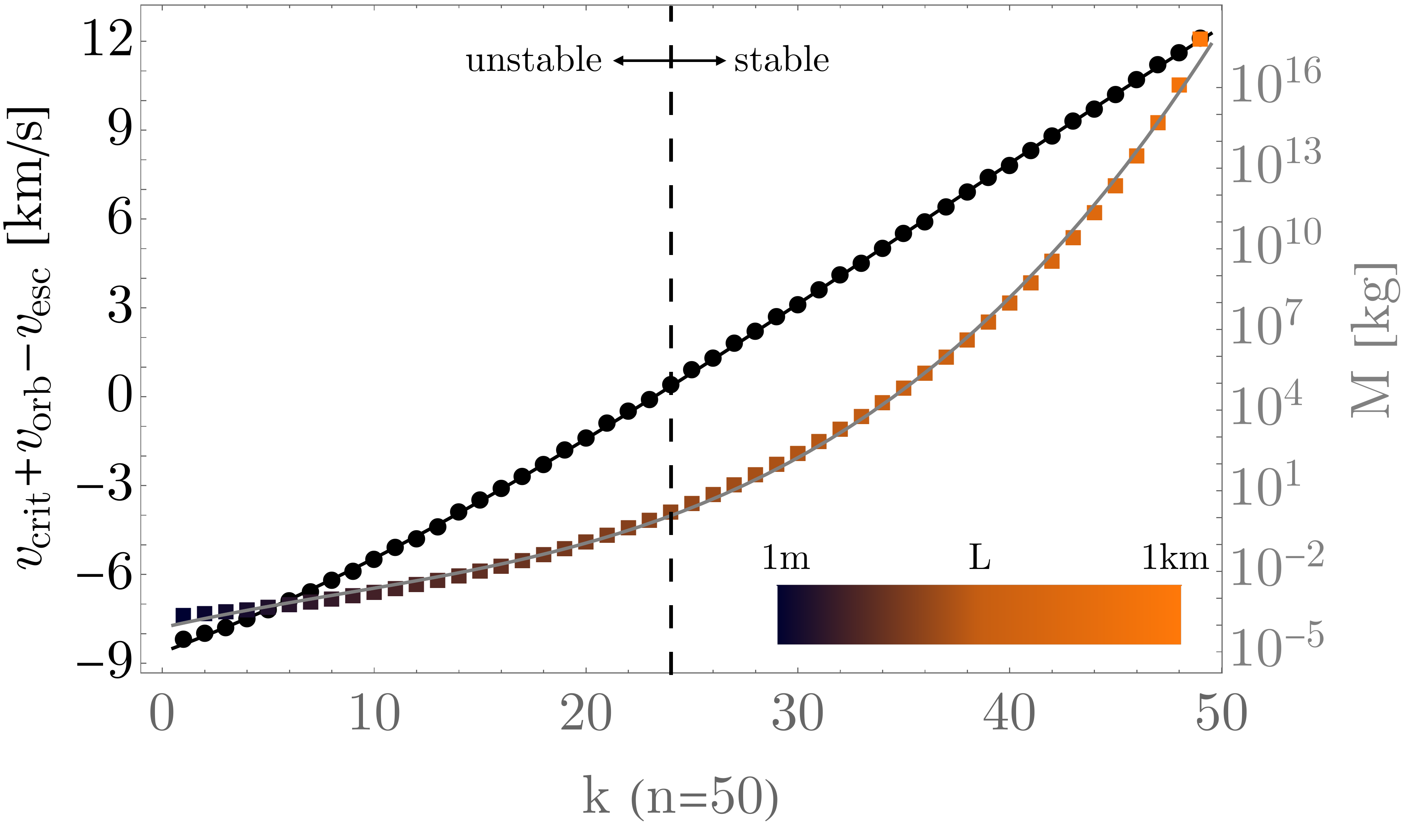}
\caption{
Black circles and black line depicts the payload's velocity at release minus
the escape velocity from our solar system (left $y$-axis) as a function of $k$,
a design parameter for the tapered ribbon design. Rust-coloured squares and grey
line depicts \catapult's mass as a function of $k$ (right $y$-axis), where one
can see the rapid ballooning of scale as $k$ increases. The colours denote the
length of the system. The vertical line delineates the region of stability
versus non-stability.
}
\label{fig:n50}
\end{center}
\end{figure}

To optimize the system, consider increasing to $n=50$, providing more fine
control over the location of the sail component. Exploring from $k=1$ to
$k=49$, it was found that stable configurations occur for $k\geq18$ and 
interstellar-capable configurations for $k\geq24$, as shown in 
Figure~\ref{fig:n50}. At large $n$, such as that used here, \catapult\ shows
enormous dynamic range in scale as a function of $k$, spanning 21
orders-of-magnitude in mass, 18 orders-of-magnitude in $W_1$ and nearly 4
orders-of-magnitude in length. Large $k$ choices lead to unreasonably large
systems and thus it is desirable to select the smallest $k$ which is both
stable and interstellar, here $k=24$.

The $n=50$ and $k=24$ design has $W_1=62.7$\,m with a payload thickness 
of 2860\,nm, which should be sufficient for a thin computer chip. The
system achieves $v_{\mathrm{crit}}=12.1$\,km/s, 
$v_{\mathrm{orb}}+v_{\mathrm{crit}}=40.4$\,km/s and has a Sun escape velocity 
of $40.0$\,km/s. We estimate a charge time of
$\omega_{\mathrm{crit}}/\dot{\omega} = 351.1$\,days. 
With a total mass of a 1.6\,kg and 
being constructed from materials already in widespread production, this
provides a possible example of a realisable interstellar sail.

Notably, the sail component here has mean density of $14.6$\,g/m$^2$ 
(total area is 164.4\,m$^2$), which 
is an order-of-magnitude heavier than a statite sail that would balance
radiation pressure with gravitational acceleration \citep{forward:1993}. Thus,
\catapult\ could achieve interstellar flight using existing materials, solar
radiation pressure only and with an thickness an order-of-magnitude greater
than that of \catapult-free solar sail. This highlights the advantages of the
system as a possible stepping stone technology to future sails.

\section{Discussion}
\label{sec:discussion}

This work has explored the possibility of using the Sun's radiation to spin-up
a thin structure in space with the ultimate goal of releasing a payload at
high velocities. Although much of this work has set the goal of reaching escape
velocity from our solar system, such a system is of course useful for
interplanetary missions too.

To our knowledge, this idea has not been previously explored in the literature
and thus much of this paper derives the basic equations governing the flux,
forces, orbit, charging time, critical velocity and design constraints. For
the sake of conciseness, this paper is not intended as an exhaustive examination
of the technical feasibility of such a system, nor is it claimed that it is
definitively cost-effective, practical or feasible - although this analysis
finds no clear objections.

With existing materials, a 1-2\,kg \catapult\ can eject smart-phone sized
sails into interstellar space with less than a year's worth of charge-up
time - powered by the Sun. The final velocities are by no means relativistic
and thus our analysis suggests such a system cannot satisfy the goals of
\textit{Breakthrough Starshot} \citep{parkin:2018}, who seek $0.2c$ (although
in what follows we discuss some mechanisms of increasing the final speed). The
system perhaps best lends itself to cases where such high speeds are not
required and, be it for cost-driven or legal-constraints, the use of ${\sim}$GW
lasers is prohibitive. The system is also in principle multi-use, as discussed
in Section~\ref{sec:backreaction}. Ultimately, this paper is an intellectual
exercise - it is the author's belief that the landscape
of possible solutions should be investigated and publicly disseminated to
provide us with the context of how best to proceed.

\subsection{Going Faster}

Although \catapult\ can achieve escape velocity from our solar system using the
Sun's radiation pressure alone, the ejection speeds are comparatively slow and
thus it is worth discussing methods by which greater speeds could be attained.
The previously discussed Oberth effect is an example of this (see
Section~\ref{sec:orbit}), but may not be cost-effective given the need to first
transfer into an eccentric Sun-grazing orbit.

An obvious improvement would be introduced by increasing the specific strength
of the materials used. In particular, the promise of large, high-grade graphene
sheets would lead to significant improvements. For example, setting $n=20$ and
$k=12$ for a $5$\,$\mu$m thick system with a 0.1m x 0.1m sail, we find that an
$L=8.2$\,m graphene \catapult\ ($\sigma_{\mathrm{bulk}}=120$\,GPa) would reach
$v_{\mathrm{crit}}+v_{\mathrm{orb}}-v_{\mathrm{esc}}=9.4$\,km/s after 867\,days
of charging (and would be stable). Briefer charge times would be possible by
modifying the orbit, as discussed in Section~\ref{sec:spinup}.

Another improvement would be to perform the same trick as used by previous
interstellar probes i.e. gravity assists. Velocity boosts of up to
${\sim}$10\,km/s are routinely possible, which can be stacked for planetary
alignments to produce significant extra speed \citep{dodd:2020}.

A third possibility is to treat \catapult\ as merely an initial launch
phase followed by acceleration with laser pressure \citep{marx:1966,
redding:1967,forward:1984}. In many ways, this echoes the approach of
\textit{SpinLaunch} - a private company attempting to use a centrifuge to
deliver 1-2\,km/s of initial launch speed followed by conventional chemical
rocket systems to escape the Earth's gravity well \citep{niederstrasser:2022}.

A fourth approach is the idea of artificially increasing the tensile strength
of the material via the use of a restoring force generated by the vehicle
itself, which is briefly discussed in the next subsection.

\subsection{Charged Paddles}

\catapult\ is ultimately limited by the tensile strength of known
materials. One possible way to increase $v_{\mathrm{crit}}$ further would be
to apply an equal and opposite electrostatic charge to each tip, $q$. In
what follows, a brief and approximate estimation of the impact of this
is explored. The opposite charges would lead to a rotating dipole of magnitude

\begin{align}
m = \frac{1}{2} q L^2 \omega.
\end{align}

A rotating magnetic dipole generates a magnetic field, which in principle
could be useful for other purposes such as a space weather protection or
controlling plasmas. A rotating dipole radiates electromagnetic waves with a
power given by

\begin{align}
P_{\mathrm{rad}} &= \frac{\mu_0 \omega^4 m^2}{6\pi c^3},\nonumber\\
&= \frac{\mu_0 q^2 L^4 \omega^6}{24 \pi c^3}.
\end{align}

The total power incident upon \catapult\ is given by 
Equation~(\ref{eqn:avgthermalpower}). Equating these two, the maximum charge
that could be used before the system would bleed power is

\begin{align}
q^2 &= \frac{48\pi c^3 A S (2-R_{\alpha}-R_{\beta})}{\pi \mu_0 L^4 \omega^6}
\end{align}

The electrostatic force bringing these charges together will thus be

\begin{align}
F &= \frac{1}{4\pi \epsilon_0} \frac{q^2}{L^2},\nonumber\\
\qquad&= \frac{12 c^5 A S (2-R_{\alpha}-R_{\beta})}{\pi\omega^6L^6}.
\end{align}

This competes with the centrifugal force of the rotating system. Equating
to that force and re-arranging for $v = \omega (L/2)$ yields

\begin{align}
v &= \Big(\frac{3 c^5}{32\pi} \frac{S L}{\Sigma}\Big)^{1/8},\nonumber\\
\qquad&\simeq 1000\,\mathrm{km}/\mathrm{s}\,\Big(\frac{S}{S_{\oplus}}\Big)^{1/8} \Big(\frac{L}{100\,\mathrm{m}}\Big)^{1/8} \Big( \frac{\Sigma}{10\,\mathrm{g}/\mathrm{m}^2} \Big)^{-1/8}
\end{align}

where it has been assumed that $R_{\alpha}=1$ and $R_{\beta}=0$ for simplicity.
Whilst certainly not comparable to the goal of \textit{Breakthrough Starshot's}
$0.2c$, a speed of 0.3\% the speed of light would quite an improvement from
previous systems, reaching Proxima Centauri in just over a millennia.




\section*{Acknowledgements}
Special thanks to donors to the Cool Worlds Lab, without whom this kind of research would not be possible:
Douglas Daughaday,
Elena West,
Tristan Zajonc,
Alex de Vaal,
Mark Elliott,
Stephen Lee,
Zachary Danielson,
Chad Souter,
Marcus Gillette,
Tina Jeffcoat,
Jason Rockett,
Tom Donkin,
Andrew Schoen,
Reza Ramezankhani,
Steven Marks,
Nicholas Gebben,
Mike Hedlund,
Leigh Deacon,
Ryan Provost,
Nicholas De Haan,
Emerson Garland,
The Queen Road Foundation Inc,
Scott Thayer,
Frank Blood,
Ieuan Williams,
Xinyu Yao \&
Axel Nimmerjahn.


\begin{thebibliography}{99}
\bibitem[\protect\citeauthoryear{Atwater et al.}{2018}]{atwater:2018}
Atwater H.~A., Davoyan A.~R., Ilic O., Jariwala D., Sherrott M.~C., Went C.~M., Whitney W.~S., et al., 2018, NatMa, 17, 861. doi:10.1038/s41563-018-0075-8
\bibitem[\protect\citeauthoryear{Benford \& Benford}{2003}]{benford:2003}
Benford J., Benford G., 2003, AIPC, 664, 303. doi:10.1063/1.1582119
\bibitem[\protect\citeauthoryear{Blanco \& Mungan}{2019}]{blanco:2019}
Blanco P.~R., Mungan C.~E., 2019, PhTea, 57, 439. doi:10.1119/1.5126818
\bibitem[\protect\citeauthoryear{Dodd}{2020}]{dodd:2020}
Dodd S., 2020, AGUFMSH01, 2020, SH019-01
\bibitem[\protect\citeauthoryear{Eastwood et al.}{2015}]{eastwood:2015}
Eastwood J.~P., Kataria D.~O., McInnes C.~R., Barnes N.~C., Mulligan P., 2015, Wthr, 70, 27. doi:10.1002/wea.2438
\bibitem[\protect\citeauthoryear{Forward}{1984}]{forward:1984}
Forward R.~L., 1984, JSpRo, 21, 187. doi:10.2514/3.8632
\bibitem[\protect\citeauthoryear{Forward}{1993}]{forward:1993}
Forward R.~L., 1993, ``Statite: Spacecraft That Utilizes Light Pressure and Method of Use'',
US patent 5183225
\bibitem[\protect\citeauthoryear{Goldstein}{1980}]{goldstein:1980}
Goldstein H., 1980, Classical Mechanics, 2nd edn. Addison-Wesley, Reading, MA
\bibitem[\protect\citeauthoryear{Jin et al.}{2022}]{jin:2022}
Jin W., Li W., Khandekar C., Orenstein M., Fan S., 2022, arXiv, arXiv:2206.05383
\bibitem[\protect\citeauthoryear{Katz}{2021}]{katz:2021}
Katz J.~I., 2021, RNAAS, 5, 61. doi:10.3847/2515-5172/abf124
\bibitem[\protect\citeauthoryear{Kezerashvili \& V{\'a}zquez-Poritz}{2009}]{keze:2009}
Kezerashvili R.~Y., V{\'a}zquez-Poritz J.~F., 2009, PhLB, 675, 18. doi:10.1016/j.physletb.2009.03.058
\bibitem[\protect\citeauthoryear{Kipping}{2019}]{quasite:2019}
Kipping D., 2019, RNAAS, 3, 91. doi:10.3847/2515-5172/ab2fdb
\bibitem[\protect\citeauthoryear{Kuzminova et al.}{2019}]{kuzminova:2019}
Kuzminova, K., Solar, P., Kus, P., Kylian, O., 2019, Journal of Nanomaterials, 2019, 1. doi:10.1155/2019/1592621
\bibitem[\protect\citeauthoryear{Lebedev}{1901}]{lebedev:1901}
Lebedev P., 1901, ``Untersuchungen über die Druckkräfte des Lichtes'', Annalen der Physik, Series 4 6, 433
\bibitem[\protect\citeauthoryear{Lubin}{2016}]{lubin:2016}
Lubin P., 2016, JBIS, 69, 40
\bibitem[\protect\citeauthoryear{Manchester \& Loeb}{2017}]{manchester:2017}
Manchester Z., Loeb A., 2017, ApJL, 837, L20. doi:10.3847/2041-8213/aa619b
\bibitem[\protect\citeauthoryear{Marx}{1966}]{marx:1966}
Marx G., 1966, Natur, 211, 22. doi:10.1038/211022a0
\bibitem[\protect\citeauthoryear{Mattox}{1998}]{mattox:1998}
Mattox D.~M., 1998, Handbook of Physical Vapor Deposition (PVD) Processing, 1st edn. William Andrew Publishing, Norwich, NY.
\bibitem[\protect\citeauthoryear{M{\'e}ndez \& Rivera-Valent{\'\i}n}{2017}]{mendez:2017}
M{\'e}ndez A., Rivera-Valent{\'\i}n E.~G., 2017, ApJL, 837, L1. doi:10.3847/2041-8213/aa5f13
\bibitem[\protect\citeauthoryear{Moeckel}{1972}]{moeckel:1972}
Moeckel W.~E., 1972, JSpRo, 9, 942. doi:10.2514/3.30415
\bibitem[\protect\citeauthoryear{Nichols \& Hull}{1903}]{nichols:1903}
Nichols E.~F., Hull G.~F., 1903, ApJ, 17, 315. doi:10.1086/141035
\bibitem[\protect\citeauthoryear{Niederstrasser}{2022}]{niederstrasser:2022}
Niederstrasser, C.G., 2022. The small launch vehicle survey: A 2021 update (The rockets are flying). Journal of Space Safety Engineering, 9(3), 341-354. doi:10.1016/j.jsse.2022.07.003
\bibitem[\protect\citeauthoryear{Ohring}{2002}]{ohring:2002}
Ohring M., 2002, Materials Science of Thin Films: Deposition and Structure, 2nd edn. Academic Press, San Diego, CA.
\bibitem[\protect\citeauthoryear{Parkin}{2018}]{parkin:2018} Parkin K.~L.~G., 2018, AcAau, 152, 370. doi:10.1016/j.actaastro.2018.08.035
\bibitem[\protect\citeauthoryear{Patsalas et al.}{2015}]{patsalas:2015}
Patsalas P., Kalfagiannis N., Kassavetis S., 2015, Materials, 8, 3128. doi:10.3390/ma8063128
\bibitem[\protect\citeauthoryear{Powers \& Coverstone}{2001}]{powers:2001}
Powers R.~B., Coverstone V.~L., 2001, JAnSc, 49, 269. doi:10.1007/BF03546322
\bibitem[\protect\citeauthoryear{Rafat et al.}{2022}]{rafat:2022}
Rafat M.~Z., Dullin H.~R., Kuhlmey B.~T., Tuniz A., Luo H., Roy D., Skinner S., et al., 2022, PhRvP, 17, 024016. doi:10.1103/PhysRevApplied.17.024016
\bibitem[\protect\citeauthoryear{Redding}{1967}]{redding:1967}
Redding J.~L., 1967, Natur, 213, 588. doi:10.1038/213588a0
\bibitem[\protect\citeauthoryear{Singal}{2017}]{singal:2017}
Singal A.~K., 2017, arXiv, arXiv:1708.05062. doi:10.48550/arXiv.1708.05062
\bibitem[\protect\citeauthoryear{Spencer et al.}{2021}]{spencer:2021}
Spencer D.~A., Betts B., Bellardo J.~M., Diaz A., Plante B., Mansell J.~R., 2021, AdSpR, 67, 2878. doi:10.1016/j.asr.2020.06.029
\bibitem[\protect\citeauthoryear{Srivastava, Chu, \& Swartzlander}{2019}]{srivastava:2019}
Srivastava P.~R., Chu Y.-J.~L., Swartzlander G.~A., 2019, OptL, 44, 3082. doi:10.1364/OL.44.003082
\bibitem[\protect\citeauthoryear{Starbuck \& Hansen}{2009}]{starbuck:2009}
Starbuck, J.~M. \& Hansen, J.~G., 2009, Office of Scientific and Technical Information (OSTI), 975070. doi:10.2172/975070
\bibitem[\protect\citeauthoryear{Tsiolkovsky}{1968}]{tsiolkovsky:1968}
Tsiolkovsky K.~E., 1968, K.~E. Tsiolkovsky Selected Works, compiled by
V.~N. Sokolosky, MIR Publishers (Moscow), English Translation
\bibitem[\protect\citeauthoryear{Weiss et al.}{1979}]{weiss:1979}
Weiss R. F., Pirri, A. N., \& Kemp, N. H. 1979, Astronautics and Aeronautics, 17, 50
\bibitem[\protect\citeauthoryear{Worden et al.}{2021}]{worden:2021}
Worden S.~P., Green W.~A., Schalkwyk J., Parkin K., Fugate R.~Q., 2021, ApOpt, 60, H20. doi:10.1364/AO.435858
\bibitem[\protect\citeauthoryear{Worrall}{1982}]{worrall:1982}
Worrall J., 1982, SHPSA, 13, 133. doi:10.1016/0039-3681(82)90023-1
\bibitem[\protect\citeauthoryear{Zander}{1964}]{zander:1964}
Zander F. 1964, Problems of flight by jet propulsion:
Interplanetary flights, NASA Technical Translation F-147
\end{thebibliography}
\bibliographystyle{aasjournalv7}



\end{document}